\documentclass[prb,twocolumn,showpacs,preprintnumbers,amsmath,amssymb]{revtex4}

\usepackage{graphicx,dcolumn,bm,SIunits,amsmath}

\begin{document}

\preprint{APS/123-QED}

\title{Frequency dependence of the radiative decay rate of excitons in self-assembled quantum dots: experiment and theory}

\author{S{\o}ren Stobbe}
\email{ssto@fotonik.dtu.dk}
\author{Jeppe Johansen}
\author{Philip Tr{\o}st Kristensen}
\author{J{\o}rn M. Hvam}
\author{Peter Lodahl}
\email{pelo@fotonik.dtu.dk}
\homepage{http://www.fotonik.dtu.dk/quantumphotonics}
\affiliation{DTU Fotonik, Department of Photonics Engineering, Technical
University of Denmark, Building 343, DK-2800 Kgs. Lyngby, Denmark}
\date{\today}

\begin{abstract}
We analyze time-resolved spontaneous emission from excitons confined
in self-assembled $\mathrm{InAs}$ quantum dots placed at various
distances to a semiconductor-air interface. The modification of
the local density of optical states due to the proximity of the interface enables unambiguous determination of the radiative and
non-radiative decay rates of the excitons. From measurements at
various emission energies we obtain the frequency dependence of
the radiative decay rate, which is only revealed due to the
separation of the radiative and non-radiative parts. It contains
detailed information about the dependence of the exciton
wavefunction on quantum dot size. The experimental results are
compared to the quantum optics theory of a solid state emitter in
an inhomogeneous environment. Using this model, we extract the
frequency dependence of the overlap between the electron and hole
wavefunctions. We furthermore discuss three models of quantum dot
strain and compare the measured wavefunction overlap to these
models. The observed frequency dependence of the wavefunction
overlap can be understood qualitatively in terms of the different
compressibility of electrons and holes originating from their
different effective masses.
\end{abstract}


\maketitle

\section{Introduction\label{sec:introduction}}

Semiconductor quantum dots (QDs) are nanoscale solid state
structures that provide three-dimensional quantum confinement of
otherwise mobile charge carriers. Self-assembled QDs of
$\mathrm{InAs}$ embedded in $\mathrm{GaAs}$ provide confinement
for both electrons and holes in a direct band gap semiconductor.
Hence, they are optically active with the benefits of a high
quantum efficiency and compatibility with existing semiconductor
technology. These properties make the QDs highly promising light
sources for novel optical devices including optical quantum
information devices \cite{Imamoglu1999}. This has led to an
increasing interest in the quantum-optical properties of QDs, and
major achievements include the demonstration of the Purcell effect
for QDs in solid-state cavities \cite{Gerard1998} and strong
coupling between a single QD and the optical mode of a cavity
\cite{Reithmaier2004,Yoshie2004}. Very recently also electrical
tuning of such quantum photonics devices was demonstrated
\cite{Kistner2008,Laucht2008}, which is a significant milestone
towards practical all-solid state cavity quantum electrodynamics
devices.

Despite the recent progress, a thorough understanding of the
dynamics of light-matter coupling for QDs in nanostructured
photonic media is still lacking. Such an understanding is required
for quantitative comparisons between experiment and theory. The
problem is two-fold, i.e. a detailed understanding of both the
optical part and the electronic part is required. The optical part
is described by the local density of optical states (LDOS)
expressing the distribution of modes that the QD can radiate to,
while the electronic part is determined by the exciton
wavefunction for the QD. Here we will investigate an optical
system where the LDOS can be calculated exactly, and use that to
extract detailed information about the QD. Our experimental
results are compared to a theoretical QD model, and the effect of
QD size, material composition, and strain is investigated. Such
quantitative comparisons of experimental data to simple
theoretical QD models are much needed in order to assess the full
potential of QDs in nanostructured media for, e.g., single-photon
sources \cite{Michler2000}, low-threshold lasers
\cite{Reitzenstein2008}, or spontaneous emission control
\cite{Lodahl2004,Julsgaard2008}.

When interpreting spontaneous emission decay curves from QDs, it
is often implicitly assumed that the QDs primarily decay through
radiative recombination, while non-radiative processes are
negligible. Unfortunately, this assumption is not generally valid,
and omnipresent non-radiative processes must be considered. Only
few experiments have addressed this issue. Robert et al.
established an upper bound on the contribution from nonradiative
processes of 25\% by measuring the ratio of the bi-exciton to
exciton emission intensity at saturation \cite{Robert2002}.
Quantitative measurements of the radiative and non-radiative decay
rates of QDs were only carried out recently using a modified LDOS
both for colloidal QDs \cite{Brokmann2004,Leistikow2009} and for self-assembled QDs \cite{Johansen2008}. Precise
measurements of the radiative decay rates are essential since
nanophotonic devices rely on the ability to manipulate the
radiative processes, while non-radiative recombination leads to
loss in the system.

As first pointed out by Purcell \cite{Purcell1946} the radiative
decay rate of an emitter is modified inside a structured
dielectric medium, which is due to the modification of the LDOS.
In early experiments by Drexhage \cite{Drexhage1970}, this effect
was experimentally demonstrated by positioning emitters in the
proximity of a reflecting surface. We have recently employed the
modified LDOS near a semiconductor-air interface as a
spectroscopic tool to extract radiative and non-radiative decay
rates and from that infer the overlap between the electron and
hole wavefunctions \cite{Johansen2008}. This technique relies on
the fact that the radiative decay rate is proportional to the
LDOS, while the non-radiative decay rate is unaffected. In the
present paper we expand on our previous work in particular by
comparing the measured radiative decay rate to theory, which
requires a detailed model of the QD electron and hole
wavefunctions. We have measured the radiative decay rate at
different emission energies, which reveals the dependence of the
QD optical properties on its size. We review the Wigner-Weisskopf
theory of spontaneous emission from QDs, predicting an exponential
decay of the exciton population and the LDOS is derived for the
applied interface geometry using a Green's function technique. We
furthermore show that the radiative decay rate of a QD in a
homogeneous medium is proportional to the square of the overlap
between electron and hole wavefunctions and calculate the
frequency dependence of this overlap using a simple two-band model
of the QD. The QD model is discussed in details and compared to
our experimental data employing realistic parameters as input to
the theory. The pronounced size dependence of the electron-hole
wavefunction overlap is found to originate from the differences in
effective mass and binding energy of the electron and hole.
Furthermore, we investigate three different strain models for the
QD and compare their predictions to experiment thereby providing
valuable insight on the complex strain mechanisms of
self-assembled QDs.

This paper is organized as follows: In section \ref{sec:method} we
present the experimental method and in section
\ref{sec:measurements} the experimental results. In section
\ref{sec:theory} we discuss the Wigner-Weisskopf model for
spontaneous emission and derive the relation between the radiative
decay rate, the LDOS, and the wavefunction overlap. In section
\ref{sec:NumericalResults} we combine the analytical expressions
for the radiative rate with the numerical results for the
wavefunction overlap and compare theory with experiment. Finally,
we present conclusions in section \ref{sec:conclusion}.

\section{Experimental technique for determining the radiative decay rate of quantum dots\label{sec:method}}

\begin{figure}
\includegraphics[width=\columnwidth]{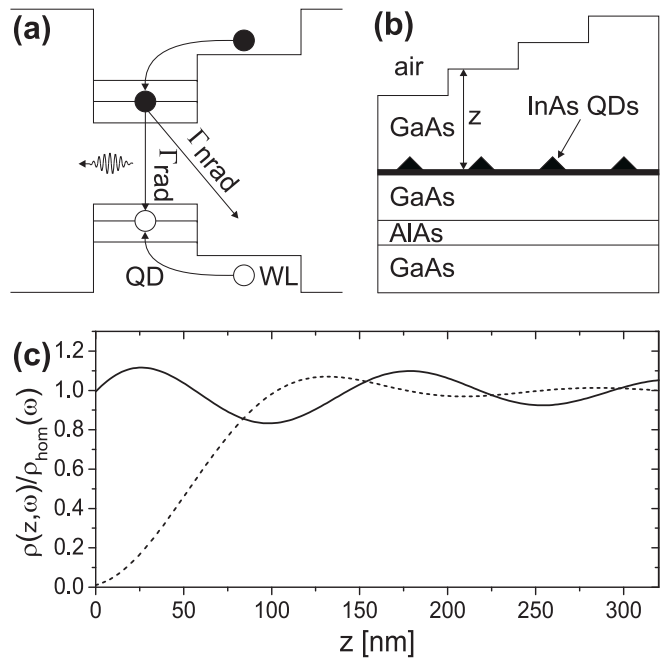}
\caption{\label{Figure1_ver17} (a) Schematic band diagram
illustrating the spontaneous emission process in a QD. An electron
is excited optically from a valence band to a conduction band
wetting layer (WL) state and the generated electron and hole relax
to the lowest energy QD state on a picosecond timescale. The
electron can subsequently decay by either radiative or
non-radiative recombination with rates $\Gamma_{rad}$ and
$\Gamma_{nrad}$ respectively. (b) Schematic illustration of the
sample under investigation. $\mathrm{InAs}$ QDs are embedded in
$\mathrm{GaAs}$ and positioned at different distances $z$ to the
$\mathrm{GaAs}$-air interface. (c) The LDOS as a function of
distance $z$ to a $\mathrm{GaAs}$-air interface for a dipole
orientation parallel (solid curve) or perpendicular (dashed curve)
to the interface.}
\end{figure}

Spontaneous emission of a photon from a QD occurs when an
electron-hole pair (an exciton) recombines, as illustrated
schematically in Fig.~\ref{Figure1_ver17}(a).  As will be shown
rigorously in section \ref{sec:theory}, the QD radiative decay
rate $\Gamma_{rad}(\mathbf{r},\omega,{\mathbf{e_p}})$ in a
structured environment is proportional to the projected local
density of optical states (LDOS)
$\rho(\mathbf{r},\omega,{\mathbf{e_p}})$, where the projection is
along the direction ${\mathbf{e_p}}$ of the transition momentum matrix element, which corresponds to the
orientation of the transition dipole moment. The LDOS is
modified in an inhomogeneous dielectric medium due to optical
reflections at interfaces. In emission experiments, the total
decay rate is measured, which can be expressed as
\cite{NanoOpticsBook}
\begin{align}
\Gamma(\mathbf{r},\omega,{\mathbf{e_p}}) =
\Gamma_{rad}^{hom}(\omega)\frac{\rho(\mathbf{r},\omega,{\mathbf{e_p}})}{\rho_{hom}(\omega)}
+ \Gamma_{nrad}(\omega),\label{eq_gamma_tot}
\end{align}
where $\rho_{hom}(\omega)$ is the density of optical states for a
homogeneous medium, and $\Gamma_{nrad}(\omega)$ is the rate for
non-radiative recombination. $\omega$ is the emission frequency
and thus $\hbar \omega$ the emission energy, and ${\bf r}$ the
position of the QD. Non-radiative recombination is due to intrinsic
QD processes and thus independent of the LDOS. $\Gamma_{rad}^{hom}(\omega)$ is the
radiative rate that the QD would exhibit in a homogenous medium
without any boundaries. In our case, the refractive index of the
medium is $n=3.5$ corresponding to that of $\mathrm{GaAs}$.
Investigating $\Gamma_{rad}^{hom}(\omega)$ in detail provides
valuable insight into the properties of the exciton wavefunction
confined in the QD potential.

The exact nature of non-radiative recombination in QDs is not yet
fully understood. It is often implicitly assumed that
non-radiative recombination is negligible, but as we will see in
the following even for very weak excitation intensities this is
not a valid assumption. Possible non-radiative processes include
surface recombination at the interfaces between the QD and the
surrounding semiconductor material, Auger processes, and trapping
of electron and/or holes at defects \cite{Chuang}, and any first
principles calculation of these effects is a tremendous task.
Reliable ways of extracting the radiative and non-radiative parts
of the decay rate are therefore essential.

The radiative and non-radiative decay rates can be separated by
time-resolved spontaneous emission measurements if the QDs are
placed in an environment with a known LDOS, cf.
Eq.~(\ref{eq_gamma_tot}).  A planar interface between two regions
with different refractive indices is the most simple example of
such an inhomogeneous dielectric medium \cite{Chance1978}. For
this particular geometry, the LDOS can be calculated exactly and
without any free parameters. Here we employ the interface between
GaAs ($n=3.5$) and air ($n=1$) as illustrated in
Fig.~\ref{Figure1_ver17}(b). We calculate the LDOS by a Green's
function technique and the results for dipole orientations
parallel or perpendicular to the interface are shown in
Fig.~\ref{Figure1_ver17}(c). We stress that no assumptions need to
be made about, e.g., the QD density and the excitation beam
profile in order to employ this experimental technique, as opposed
to alternative ways of determining the radiative decay rate such
as by absorption spectroscopy
\cite{Birkedal2000,Warburton1997,Silverman2003}.

\section{Measurements of spontaneous emission decay rates near a semiconductor-air interface\label{sec:measurements}}
\begin{figure}
\includegraphics[width=\columnwidth]{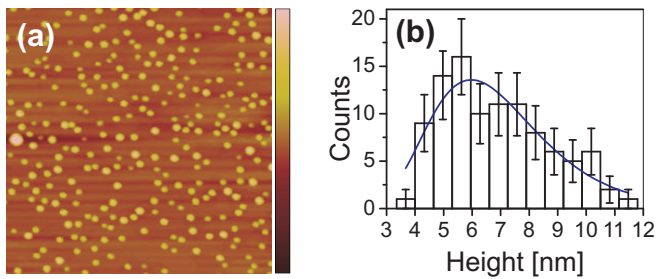}
\caption{(Color online) (a) Topographic atomic force
micrograph depicting a surface area of $1\times 1$~$\mathrm{\micro
m}$ of uncapped QDs on the unprocessed wafer. The color scale runs
from 0~$\mathrm{nm}$ to 20~$\mathrm{nm}$. (b) Histogram of the QD
height measured by analysis of the AFM data. The blue line is a
fit to the histogram data using a log-normal distribution as
discussed in the text.\label{Figure2_ver17}}
\end{figure}

\begin{figure}
\includegraphics[width=\columnwidth]{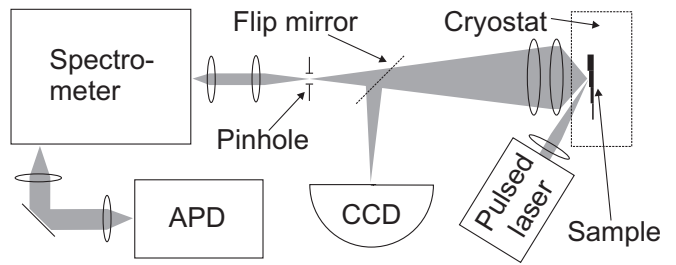}
\caption{Illustration of the optical
measurement setup. The sample is kept at 14 $\mathrm{K}$ in a
cryostat and illuminated by a pulsed laser. The spontaneously
emitted light is collected and can be directed either to a CCD
camera for sample alignment or to a spectrometer equipped with a
fast single photon counting avalanche photodiode (APD) for
time-resolved measurements.\label{Figure3_ver17}}
\end{figure}

\begin{figure}
\includegraphics[width=\columnwidth]{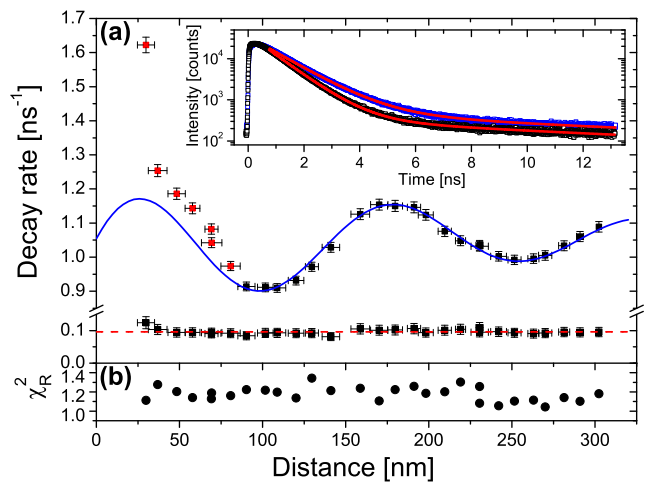}
\caption{(Color online) (a) Measured fast decay rate at an emission energy of
1.204~$\mathrm{eV}$ versus the distance to the interface (black
and red squares, upper trace). The blue line shows a theoretical
fit to the data using the exact expression for the LDOS and
Eq.~(\ref{eq_gamma_tot}) for a dipole orientation parallel to the
semiconductor-air interface. The data points marked in red are
excluded from the fit. The lower trace shows the slow decay rate
which does not depend on the distance to the interface but retains
a constant value given by the dashed red horizontal line. Inset:
Decay curves measured for two identical QD ensembles at two
different distances to the interface at an emission energy of
$1.204$~$\mathrm{eV}$. The data shown with blue squares (black
triangles) are obtained for $z = \mathrm{109}$~$\mathrm{nm}$ ($z =
\mathrm{170}$~$\mathrm{nm}$). The red lines are fits to the data
using a bi-exponential model and we extract the decay rates from
these fits. (b) Values of the goodness-of-fit parameter $\chi_r^2$
obtained for the individual bi-exponential fits. The values
obtained are very close to the ideal value of unity.
}\label{Figure4_ver17}
\end{figure}

\begin{figure}
\includegraphics[width=\columnwidth]{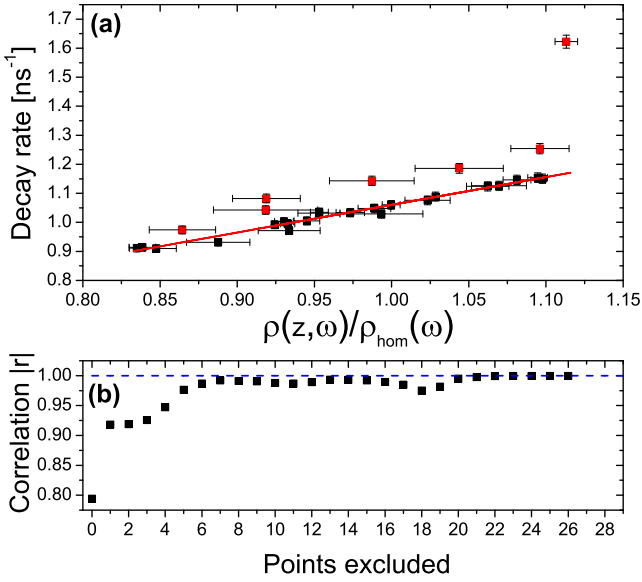}
\caption{(Color online) (a) Measured decay rate at an emission energy of 1.204
$\mathrm{eV}$ versus the normalized LDOS (black and red squares)
for a dipole orientation parallel to the semiconductor-air
interface. The red line shows a linear fit where only the black
points have been included. The red points have been omitted
because at these distances from the interface a systematic
deviation from theory was observed, as quantified by the data in
the lower panel. (b) Correlation between data and theory for the
linear regression when systematically excluding the points nearest
the interface in the fit. The correlation parameter converges to a
value close to unity (dashed blue line) when seven points have
been excluded.\label{Figure05_ver17}}
\end{figure}

\begin{figure}
\includegraphics[width=\columnwidth]{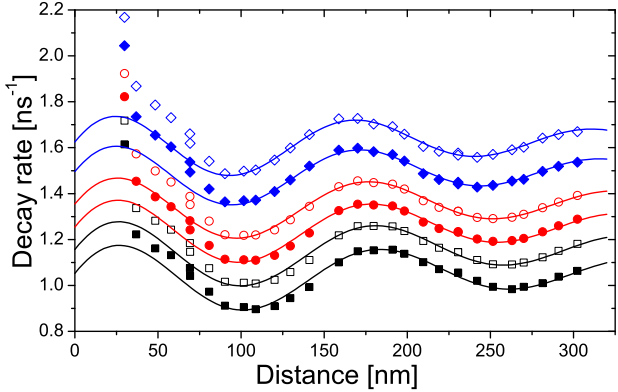}
\caption{(Color online) Decay rates as a function of distance to the interface
for six different emission energies. Each curve has been
vertically offset by $0.1$ $\mathrm{ns^{-1}}$ for visual clarity.
The fits have been obtained using a systematic exclusion of data
points near the interface, cf. Fig.~\ref{Figure05_ver17}(b). The
emission energies are 1.170 $\mathrm{eV}$ (solid black squares),
1.187 $\mathrm{eV}$ (open black squares), 1.204 $\mathrm{eV}$
(solid red circles), 1.216 $\mathrm{eV}$ (open red circles), 1.252
$\mathrm{eV}$ (solid blue diamonds), and 1.272 $\mathrm{eV}$ (open
blue diamonds). For all emission energies we note the excellent
agreement with theory for $z >
75$~$\mathrm{nm}$.\label{Figure06_ver17}}
\end{figure}

\begin{figure}
\includegraphics[width=\columnwidth]{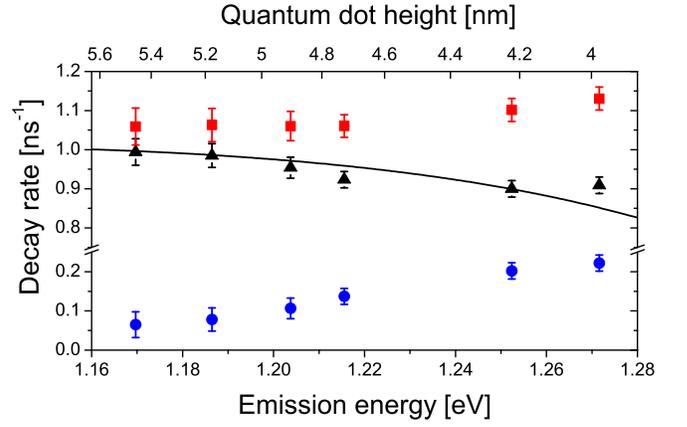}
\caption{(Color online) (a) The total decay rate (red squares), radiative decay
rate (black triangles) and non-radiative decay rate (blue circles)
as a function of QD emission energy. The radiative rate decreases
with energy but the non-radiative rate simultaneously increases so
that the total (measured) decay rate increases with increasing
energy. The solid black line is the result of the theoretical
model of the radiative decay rate using an aspect ratio of $1/2$,
omitting the wetting layer, and for an indium mole fraction of
0.95. The top scale shows the heights used in the calculation. The
details of this calculation are presented in section
\ref{sec:NumericalResults}. \label{Figure07_ver17}}
\end{figure}

\begin{figure}
\includegraphics[width=\columnwidth]{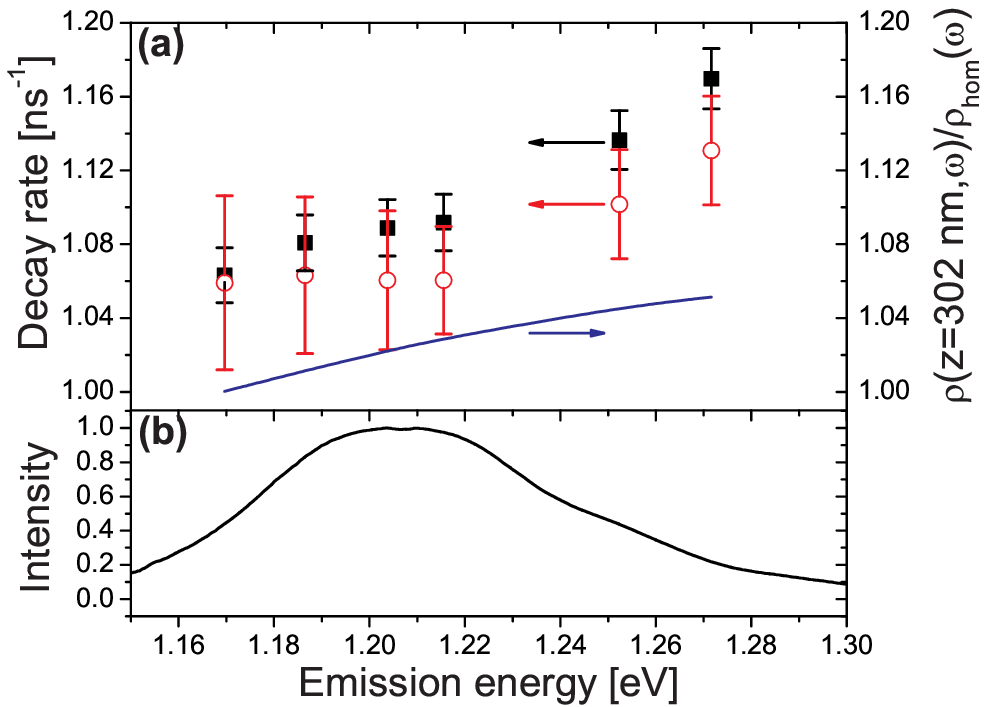}
\caption{(Color online) (a) The fast decay rate obtained from measurements on the
unprocessed wafer (solid black squares) and the total decay rate
in a homogenous medium (open red circles) obtained using the
rigorous separation of radiative and non-radiative homogeneous
decay rates. Evidently, the frequency dependence of the normalized
LDOS (solid blue line) for the unprocessed wafer
($z=302$~$\mathrm{nm}$) results in faster decay rates for QDs on
the unprocessed wafer than would be the case in a homogeneous
medium. (b) The (normalized) inhomogeneously broadened emission
spectrum of the QDs obtained under weak
excitation.}\label{Figure08_ver17}
\end{figure}

The starting point of our investigations is a $\mathrm{GaAs}$
wafer grown by molecular beam epitaxy (MBE). The QDs were grown
using the Stranski-Krastranow method on a (001) $\mathrm{GaAs}$
substrate. The growth sequence was $50$~$\mathrm{nm}$ AlAs,
$610$~$\mathrm{nm}$ $\mathrm{GaAs}$, 2.0 monolayers (MLs)
$\mathrm{InAs}$, $300$~$\mathrm{nm}$ $\mathrm{GaAs}$ cap, and
finally 2.0 ML $\mathrm{InAs}$. The AlAs layer was included for an
optional epitaxial lift-off process, which was not employed here.
Both $\mathrm{InAs}$ layers formed self-assembled QDs, but only
the embedded layer was optically active because non-radiative
surface recombination dominates for QDs at the surface. However,
since the QDs at the surface were fabricated under identical
growth conditions as the embedded ones, the density of the active
QDs can be determined by atomic force microscopy (AFM) of the
sample surface. Such an atomic force micrograph is shown in
Fig.~\ref{Figure2_ver17}(a). The image contains detailed
information about the geometry and height of the QDs at the
surface, but the topography of the surface is convolved with the
AFM tip shape function, and thus the width and exact geometry of
the QDs cannot be extracted directly. The maximum height, however,
is not subject to this effect and we used the AFM data in
Fig.~\ref{Figure2_ver17}(a) to obtain a histogram of the heights
recorded for 100 randomly selected QDs. The result is shown in
Fig.~\ref{Figure2_ver17}(b). We fit the height histogram by a
log-normal distribution given by $f(h) = \frac{h_0}{\sigma
\sqrt{2\pi} h} \exp \left( -\frac{(ln(h)-\mu)^2}{2\sigma^2}
\right)$ with $h_0 = 68$, $\sigma = 0.32$, and $\mu = 1.9$, where
$h$ is a dimensionless length scale normalized to 1~$\mathrm{nm}$. We find an average aspect
ratio defined as the height to lateral base diameter ratio of
$1/3.6$, but due to the convolution effect this is an upper bound.
Since the transition energy of a QD depends sensitively on its
height, the height distribution function is central when testing
theoretical models of the emission spectrum against experiments,
as discussed in section \ref{sec:NumericalResults}.

From the AFM data we found a QD density of 250 $\micro
\mathrm{m}^{-2},$ which corresponds to an average interdot
distance of 60~$\mathrm{nm}$. This number should be compared to
typical length scales for various relevant QD interactions.
Carrier tunneling is negligible for distances beyond
15~$\mathrm{nm}$ \cite{Wang2004} and the dipole-dipole interaction
is only significant for distances close to the F\"orster radius,
which is typically 2-9~$\mathrm{nm}$ \cite{NanoOpticsBook}. Therefore the
measurements performed here provide ensemble averaged values of
single QD properties with the advantage of an excellent
signal-to-noise ratio in the measurements.

The wafer was processed by standard UV lithography and wet
chemical etching in five subsequent steps with nominal etch depths
of $160$~$\mathrm{nm}$, $80$~$\mathrm{nm}$, $40$~$\mathrm{nm}$,
$20$~$\mathrm{nm}$, and $10$~$\mathrm{nm}$ by which we obtained 32
fields with specific distances from the QDs to the semiconductor
surface. The 32 fields were nominally equidistantly spaced with
$10$~$\mathrm{nm}$ spacing. The wet etching was done using an
etchant comprised of $\mathrm{H_3PO_4}$ (85\%), $\mathrm{H_2O_2}$
(30\%), and $\mathrm{H_2O}$ in the ratio $3:1:60$, which has an
etch rate on $\mathrm{GaAs}$ at room temperature of
1~$\mathrm{nm/s}$. We found that this etchant results in surfaces
of good optical quality with low surface roughness. A schematic
illustration of the resulting sample is shown in
Fig.~\ref{Figure1_ver17}(b). Finally we measured the actual
distance from the QDs to the semiconductor surface by using a
combination of secondary ion mass spectroscopy and surface
profiling from which we found typical depth uncertainties of
$\pm\mathrm{3.0 nm}$.

The experimental setup is illustrated in Fig.~\ref{Figure3_ver17}.
The sample was kept at $\mathrm{14 K}$ and irradiated by a
mode-locked Ti:sapphire laser emitting 300~$\mathrm{fs}$ pulses at
1.45~$\mathrm{eV}$, which corresponds to excitation of the wetting
layer states of the QD ensemble. The repetition rate was
80~$\mathrm{MHz}$ and we used an excitation intensity of
7~$\mathrm{kW/cm^2}$, which corresponds to less than 0.1~excitons
per QD generated in the wetting layer per pulse, i.e. only the QD
ground state is populated. Excitation of the WL states is
advantageous since the same density of excitons is generated
independent of sample thickness, which would not be the case for
excitation in the $\mathrm{GaAs}$ barrier since the samples have
different thicknesses. The pump configuration is illustrated in
Fig.~\ref{Figure1_ver17}(a). The spontaneous emission from the QD
ensemble was collected and then dispersed by a monochromator with
a spectral resolution of 2.6~$\mathrm{meV}$ from which it was
directed onto a fast silicon avalanche photodiode (APD).

Representative examples of two decay curves obtained for two
different distances to the interface are shown in the inset of
Fig.~\ref{Figure4_ver17}(a). We model our data with a
bi-exponential decay and typically find a goodness-of-fit
parameter $\chi_r^2$ \cite{Lakowicz} of 1.2, which is close to the
ideal value of unity, thus confirming the validity of the model.
The important parameter extracted from the fit is the fast decay
rate, which is due to recombination of bright excitons in the QD,
i.e., it equals the total decay rate discussed in the previous
section.

We measured the decay curves for 32 nominally
equidistantly spaced distances to the interface. We found that for
the two distances closest to the interface there was no detectable
spontaneous emission due to the very close proximity of the
interface and/or damage by the etching process. In
Fig.~\ref{Figure4_ver17}(a) we show the extracted fast decay rates
(upper trace) versus the measured distance from the QDs to the
interface for the remaining $30$ samples. As shown in
Fig.~\ref{Figure4_ver17}(b) we obtain values of $\chi_r^2$ near
unity for the bi-exponential fits for all distances. The solid
blue curve is a fit to the fast decay rates using the calculated
LDOS for a dipole oriented parallel to the interface. When
omitting the seven data points closest to the interface, as
discussed in detail below, we find an excellent agreement with
theory. Since the slow rate (lower trace) does not depend on the
distance to the interface and therefore does not depend on the
LDOS it must be dominated by non-radiative decay. It is attributed
as due to the recombination of dark excitons and the dynamics will
be the subject of a future publication
\cite{Johansen2008DarkExcitons}. In the remainder of this paper
we will consider only the fast decay rate.

 When plotting the fast decay rates as a function of the LDOS normalized to the density of states of a homogeneous medium,
 a linear dependence is expected, see Fig.~\ref{Figure05_ver17}(a). However, close to the interface
$(z\le 75\mathrm{nm})$ the measured decay rate is found to be systematically
larger than expected by theory. This deviation could be due to
enhanced recombination rates induced by, e.g., scattering or
impurities at the semiconductor surface. In order to exclude these
effects in our analysis of intrinsic QD properties, we
systematically excluded the data points closest to the interface
in the fit. We found that the linear regression correlation
parameter \cite{Barford} obtained from the fit in
Fig.~\ref{Figure05_ver17}(a) saturated close to the ideal value of
unity when excluding the seven innermost data points, see
Fig.~\ref{Figure05_ver17}(b). These points were consequently
abandoned in the analysis. For distances $z\ge 75\mathrm{nm}$ we
find an excellent agreement between theory and experiment, which
allows reliable extraction of the radiative and non-radiative
rates of the QDs.

The linear fit in Fig.~\ref{Figure05_ver17}(a) is based on
Eq.~(\ref{eq_gamma_tot}) and contains two free parameters, namely
the homogeneous radiative decay rate $\Gamma_{rad}^{hom}$ and the
non-radiative decay rate $\Gamma_{nrad}$. We obtain
$\Gamma_{rad}^{hom} = 0.95\pm 0.03\mathrm{ns^{-1}} $ and
$\Gamma_{nrad} = 0.11\pm 0.03\mathrm{ns^{-1}}$ at an emission
energy of $1.204\electronvolt$. For reference, the theoretical
curve for a dipole oriented perpendicular to the interface (i.e.
parallel to the growth direction) is shown in
Fig.~\ref{Figure1_ver17}(c). Clearly, this model cannot fit the
experimental data, which confirms that the QDs are oriented in the
plane perpendicular to the growth direction, which was previously
established by absorption measurements \cite{Cortez2001}. We
measured the decay curves at six energies across the
inhomogeneously broadened emission spectrum. When extracting the
fast rate from the fits, we obtain the curves shown in
Fig.~\ref{Figure06_ver17}. The different emission frequencies
shift the curves along the abscissa and more importantly the
amplitude and ordinate offsets are changing, which corresponds to
changes in $\Gamma_\mathrm{rad}$ and $\Gamma_{nrad}$. This shows
directly the frequency dependence of $\Gamma^\mathrm{hom}_{rad}$
and $\Gamma_{nrad}$, which will be discussed below.

The total decay rate in a homogeneous medium was extracted from
Fig.~\ref{Figure06_ver17} by the method outlined above and the
result is shown in Fig.~\ref{Figure07_ver17}(a). The total decay
rate increases with increasing emission energy, which could
suggest that the radiative rate increases with energy, as has been
reported for colloidal QDs \cite{vanDriel2005}. However, the opposite
turns out to be true for self-assembled QDs. In this case,
$\Gamma_{rad}^{hom}(\omega)$ is found to decrease with increasing
energy, and the overall increase in the total rate is due to the
pronounced increase in $\Gamma_{nrad}(\omega)$ with emission
energy. It should be stressed that such variations in the
radiative rate can be assessed only because a modified LDOS is
employed allowing to separate radiative and non-radiative
contributions. The striking energy dependence of the radiative
rate can be explained as being due to the dependence of the
electron and hole wavefunctions on the size of the QD, which will
be discussed and analyzed in detail in section
\ref{sec:NumericalResults}.

Here we want to stress some potential pitfalls in the interpretation of
the frequency dependence of spontaneous emission decay rates from
QDs. As already pointed out, it is decisive to include the effects
of non-radiative recombination and implement a technique that
allows to separate it from radiative recombination. Thus, an a
priori assumption of negligible non-radiative recombination would
erroneously lead to the conclusion that the radiative rate
increases with energy. Furthermore, when extracting quantitative
data for the QD decay rates, one has to be aware of the possible
influence of the presence of interfaces. In a homogeneous medium
the LDOS is proportional to the emission energy squared, but this
is not the case in proximity of interfaces. In
Fig.~\ref{Figure08_ver17}(a) we compare the measured total decay
rate versus emission energy for an unprocessed wafer and compare
to the total rate that QDs in a homogeneous medium would exhibit.
The latter has been obtained using the LDOS technique explained
above and provides "undisturbed" QD properties. For the
unprocessed wafer, the distance to the interface is
302~$\mathrm{nm},$ and for this distance the LDOS increases with
increasing energy, cf.
the solid blue line in Fig.~\ref{Figure08_ver17}(a). This
frequency dependence of the LDOS modifies the measured emission rates, which
should be taken into account, and is an example of the importance of considering
nearby interfaces in quantitative assessment of QD properties. Experimentally this issue could
be solved by growing a very thick capping layer on top of the QDs.

\section{Theoretical description of spontaneous emission from quantum dots due to interaction with the quantized electromagnetic
field\label{sec:theory}}
In this section, we give a theoretical description of the
radiative decay of excitons in QDs.
For sufficiently small QDs, the energy difference between bound
states in the QDs is much larger than the Coulomb energy and the
effect of the Coulomb interaction on the internal exciton dynamics
becomes negligible \cite{Hanamura1988}. This means that the
electron and hole comprising the exciton may be considered
independent, which is the strong confinement model. Furthermore,
we employ here a two-band description of the QD including the
effects of a wetting layer and strain, which is sufficient to
capture the essential properties of QDs
\cite{Melnik2004,Markussen2006}. Our objective here is to explore
the validity of this model by a thorough comparison to our
measurements of the radiative decay rate and the emission
spectrum, which is carried out in section
\ref{sec:NumericalResults}. Thorough explorations of even simple
QD models are much needed since complete microscopically correct
QD models are outside reach both due to the lack of experimental
knowledge about exact atomic composition and computational
complexity.

Spontaneous emission occurs due to the interaction of the exciton
with the continuum of vacuum modes. A rigorous description of
spontaneous emission requires a fully quantum description where
both the radiation field and the exciton states in the QD are
quantized. We employ here the Wigner-Weisskopf model of
spontaneous emission, which is valid when the LDOS varies slowly
with frequency over the linewidth of the QD. This is an excellent
approximation for the dielectric structures investigated here, but
may break down for QDs in photonic crystal leading to intricate
non-Markovian dynamics \cite{Kristensen2008}. The
radiative coupling strength is determined by the electron momentum
matrix element and depends furthermore sensitively on the overlap
between the electron and hole envelope wavefunctions that in turn
gives rise to the frequency dependence of the radiative decay
rate. We derive here this frequency dependence, which will be
compared in detail to the experimental data in section
\ref{sec:NumericalResults}.

\subsection{Wigner-Weisskopf theory of spontaneous emission from quantum dots}

\begin{figure}
\includegraphics[width=\columnwidth]{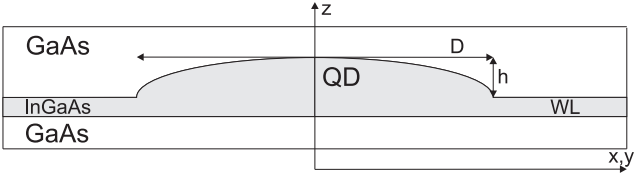}
\caption{Schematic illustration of the QD geometry used to calculate the
envelope wavefunctions. We consider a lens-shaped QD with lateral base diameter $D$ and height $h$ consisting of
$\mathrm{In_cGa_{1-c}As}$ on a wetting layer (WL). The symmetry
axis $z$ is indicated along with the radial directions
$x,y$.\label{Figure10_ver17}}
\end{figure}

According to effective mass theory, the solution to the
Schr\"odinger equation for an electron in a solid is given by
$\Psi_n(\mathbf{r}) = F_n(\mathbf{r})
u_{n,\mathbf{0}}(\mathbf{r})$, where $F_n(\mathbf{r})$ is the
envelope function, $u_{n,\mathbf{0}}(\mathbf{r})$ is the Bloch
function evaluated at the band edge $\mathbf{q} = \mathbf{0}$, and
$n \in \{c,v\}$ denotes the conduction ($c$) or valence ($v$)
band. The envelope function is the solution to the effective mass
Schr\"odinger-like equation governed by the Hamiltonian
\cite{Li1996}
\begin{align}
{H}_0(\mathbf{r}) = H_\mathrm{kin,n} + V_n(\mathbf{r})\label{eq:TheEffectiveMassHamiltonian},
\end{align}
where $H_\mathrm{kin,n}$ is the kinetic energy operator, and
$V_n(\mathbf{r})$ is the band confinement potential. Here we
consider lens-shaped QD geometries as shown in
Fig.~(\ref{Figure10_ver17}). For the conduction band the
effective mass is isotropic, so that $H_\mathrm{kin} = -
\frac{\hbar^2}{2m_0} \nabla \cdot \frac{1}{m_n(\mathbf{r})}
\nabla$, where $m_0$ is the elementary electron mass and
$m_n(\mathbf{r})$ is the effective mass. For the valence band the
anisotropy of the effective mass must be taken into account, which
is discussed in Appendices~\ref{sec:strain_appendix} and \ref{sec:NumericalModelingAppendix}.

For the valence band $m_v < 0$ and $V_n(\mathbf{r}) < 0$, which
leads to negative eigenenergies. In the electron-hole
representation \cite{Haug_Koch} we define the hole in the valence
band as a particle with positive effective mass $m_h(\mathbf{r}) =
- m_v(\mathbf{r})$ subject to a positive confinement potential
$V_h(\mathbf{r}) = - V_v(\mathbf{r})$ yielding positive
eigenenergies. Clearly, the envelope function remains the same in
the new representation, i.e. $F_h(\mathbf{r}) =
F_v(\mathbf{r})$. For III-V semiconductors the valence band is
comprised of degenerate bands with different effective masses.
However for QDs strain lifts this degeneracy and we may consider
the valence band as a single band (the heavy hole band). We
discuss details of the band structure in the presence of strain in
Appendix~\ref{sec:strain_appendix}.

We describe the light-matter interaction by the Hamiltonian
$H'(\mathbf{r},t) = -\frac{q}{m_0} \mathbf{p} \cdot
\mathbf{A}(\mathbf{r},t)$, where $q$ is the elementary charge,
$\mathbf{p}$ is the momentum operator  and
$\mathbf{A}(\mathbf{r},t)$ is the vector potential of the
quantized electromagnetic field. The latter is given by
\cite{Vats2002,Loudon}
\begin{align}
\mathbf{A}(\mathbf{r},t)= \sum_{\mathbf{\mu}}
\frac{\epsilon_\mathbf{\mu}}{\omega_\mathbf{\mu}}
\hat{\mathbf{e}}_\mathbf{\mu} \left( A_\mathbf{\mu}(\mathbf{r})
a_\mathbf{\mu} e^{-i \omega_\mathbf{\mu} t}+
A^\ast_\mathbf{\mu}(\mathbf{r}) a^\dagger_\mathbf{\mu} e^{i
\omega_\mathbf{\mu} t} \right). \label{eq_ApotentialQuantum1}
\end{align}
Here $\mathbf{\mu} = (\mathbf{k},s)$ is the combined wavevector
$\mathbf{k}$ and polarization index $s \in \{ 1,2 \}$,
$\omega_\mathbf{\mu}$ is the optical angular frequency,
$\epsilon_\mathbf{\mu} = \sqrt{\frac{\hbar \omega_\mathbf{\mu}}{2
\epsilon_0}}$ is a normalization constant with $\epsilon_0$
denoting the vacuum permittivity, $\hat{\mathbf{e}}_\mathbf{\mu}$
is a unit vector in the direction of the polarization $s$,
$A_\mathbf{\mu}(\mathbf{r})$ is the spatial field distribution
function, and $a_{\mathbf{\mu}}$ and $a_{\mathbf{\mu}}^\dagger$ are
the field annihilation and creation operators, respectively. In a homogeneous medium the field distribution functions are given by plane waves $A_\mathbf{\mu}(\mathbf{r}) = \frac{e^{i \mathbf{k} \cdot \mathbf{r}}}{\sqrt{\epsilon_r V}}$, where $\epsilon_r=n^2$ is the relative static permittivity of the material. We
will be working in the Coulomb gauge in which the scalar potential
of the electromagnetic field is zero and the divergence of the
vector potential vanishes.

We consider the initial state $|i\rangle = |c\rangle \otimes | 0
\rangle$ with $|c\rangle = F_e(\mathbf{r}) |u_{c,0}\rangle$
corresponding to a QD with one electron promoted to the conduction
band and no photons in the radiation field. The final state
relevant for spontaneous emission is $|f_\mathbf{\mu}\rangle =
|v\rangle \otimes | 1_\mathbf{\mu} \rangle$ with $|v\rangle = F_h(\mathbf{r}) |u_{v,0}\rangle$, where a photon is radiated to
the mode $\mu$ while the excited electron has decayed to the
valence band. We have deliberately not written the spatial
dependence of $| c \rangle$ and $| v \rangle$ for visual clarity. We
can expand the interaction Hamiltonian by insertion of complete
sets to obtain an expression containing the raising and lowering
operators of the electronic system, which we define as $\sigma_+ =
| c \rangle \langle v |$ and $\sigma_- = | v \rangle \langle c |$,
respectively. It is convenient to change to the interaction
picture, in which the time-evolution of the raising and lowering
operators are given by $\tilde{\sigma}_+(t) = \sigma_+ e^{i
\omega_0 t}$ and $\tilde{\sigma}_-(t) = \sigma_- e^{-i \omega_0
t}$, where we have introduced the energy of the exciton transition
$\hbar \omega_0$. Furthermore, assuming that the spatial
distribution functions are slowly varying on the scale of the
wavefunctions they can be evaluated at the center position
$\mathbf{r}_0$ of the QD, which is the dipole approximation. The
interaction Hamiltonian in the interaction picture now reads
\begin{align}
\begin{split}
H'(\mathbf{r}_0&,t) =\\
& -\frac{q}{m_0} \langle c | \mathbf{p} | v \rangle \sigma_+ \cdot \sum_{\mathbf{\mu}}
\frac{\epsilon_\mathbf{\mu}}{\omega_\mathbf{\mu}}
\hat{\mathbf{e}}_\mathbf{\mu} A_\mathbf{\mu}(\mathbf{r}_0)
a_\mathbf{\mu} e^{-i \Delta_\mathbf{\mu} t}\\
 &-\frac{q}{m_0} \langle v | \mathbf{p} | c \rangle \sigma_-
\cdot \sum_{\mathbf{\mu}}
\frac{\epsilon_\mathbf{\mu}}{\omega_\mathbf{\mu}}
\hat{\mathbf{e}}_\mathbf{\mu} A^\ast_\mathbf{\mu}(\mathbf{r}_0)
a^\dagger_\mathbf{\mu} e^{i \Delta_\mathbf{\mu}
t},\label{eq:Hamiltonian6}
\end{split}
\end{align}
where $\Delta_\mathbf{\mu} = \omega_\mathbf{\mu}-\omega_0$ and we have omitted the two terms proportional to $e^{\pm i
\Delta_\mathbf{\mu} t}$ since they are rapidly
oscillating as a function of time, which is the rotating wave
approximation.

The general state vector of the system can be expanded as
\begin{align}
|\Psi(t) \rangle = c_e(t) |i\rangle + \sum_\mu
c_\mathbf{\mu}(t)|f_\mathbf{\mu}\rangle, \label{eq:Psi}
\end{align}
and inserting into the Schr\"odinger equation in the interaction
picture leads to the following equation of motion \cite{Vats2002}
\begin{eqnarray}
\begin{split}
\frac{d}{dt} &c_e(t) =  -\frac{q^2}{2 \hbar m_0^2 \epsilon_0 }
\left| \langle v | \mathbf{p} | c \rangle \right|^2 \\& \times \int_0^t dt'
c_e(t') \int_{-\infty}^\infty d\omega
 \frac{\rho(\mathbf{r}_0,\omega,\mathbf{\hat{e}}_\mathbf{p})}{\omega}  e^{-i (\omega-\omega_0) (t-t')}  \label{eq_001}
\end{split}
\end{eqnarray}
where we have included an integration over a Dirac delta function
in frequency, and assumed that the momentum matrix element is
constant within the linewidth of the QD.
$\rho(\mathbf{r}_0,\omega,\mathbf{\hat{e}}_\mathbf{p})$ is the
projected LDOS defined as
\begin{align}
\rho(\mathbf{r}_0,\omega,\mathbf{\hat{e}}_\mathbf{p}) =
\sum_{\mathbf{\mu}} |\mathbf{\hat{e}}_\mathbf{p} \cdot
\mathbf{\hat{e}}_\mathbf{\mu}|^2
\left|A_\mathbf{\mu}(\mathbf{r}_0) \right|^2
\delta(\omega-\omega_\mathbf{\mu}),\label{eq:LDOS_def}
\end{align}
where $\mathbf{\hat{e}}_\mathbf{p}$ is the unit vector specifying
the direction of $\langle v | \mathbf{p} | c \rangle$. This direction is determined by the Bloch matrix element as discussed below.
Since $\rho(\mathbf{r}_0,\omega,\mathbf{\hat{e}}_\mathbf{p})/\omega$ in
Eq.~(\ref{eq_001}) is slowly varying over the linewidth of the
emitter so that it can be evaluated at the emission frequency $\omega_0$
and taken outside the integral. In this case the QD population
decays exponentially $|c_e(t)|^2 =
e^{-\Gamma_{\mathrm{rad}}(\mathbf{r}_0,\omega_0,\mathbf{\hat{e}}_\mathbf{p})
t}$ with the radiative decay rate given by
\begin{align}
\Gamma_\mathrm{rad}(\mathbf{r}_0,\omega_0,\mathbf{\hat{e}}_\mathbf{p})
= \frac{\pi q^2}{\hbar m_0^2 \epsilon_0} \left| \langle v |
\mathbf{p} | c \rangle \right|^2
\frac{\rho(\mathbf{r}_0,\omega_0,\mathbf{\hat{e}}_\mathbf{p})}{\omega_0}.\label{eq:Gamma_rad}
\end{align}
This is the Wigner-Weisskopf result for spontaneous emission from
solid-state emitters. It states that the radiative decay rate is
proportional to the projected LDOS and the momentum matrix
element. In the following subsection, we discuss the evaluation of
these two terms. In the experiment, the number of photons emitted
per time is measured, which is given by
\begin{align}
N(t) = \alpha \Gamma_{rad} e^{-(\Gamma_{\mathrm{rad}} +
\Gamma_{\mathrm{nrad}}) t},
\end{align}
where additionally the rate for non-radiative recombination has
been added, and $\alpha$ is an overall scaling parameter
determined by the detection efficiency and the total number of
photons recorded during the measurement period.

\subsection{Evaluation of the projected LDOS and the transition matrix element}

The projected LDOS can be calculated using a Green's function
technique. In terms of the Green's tensor $\mathbf{G}(
\mathbf{r},\mathbf{r}',\omega )$, the projected LDOS is given by
\cite{NanoOpticsBook,Dung2000}
\begin{align}
\rho(\mathbf{r},\omega,\mathbf{\hat{e}}_\mathbf{p}) = \frac{2
\omega}{\pi c_0^2}\left( \hat{\mathbf{e}}_\mathbf{p} \cdot
\mathrm{Im}\left(\mathbf{G}( \mathbf{r},\mathbf{r},\omega )
\right) \cdot \hat{\mathbf{e}}_\mathbf{p}
\right),\label{eq:LDOS_in_terms_of_G}
\end{align}
where $c_0$ is the speed of light in vacuum. The LDOS is a
classical electromagnetic quantity obtained by solving Maxwell's
equations. However, it enters the quantum optical theory of
light-matter interaction where it describes the local density of
vacuum modes that spontaneous emission can occur to. For the
particular case of a semiconductor-air interface as considered
here, the Green's tensor is obtained by solving the following
closed expression \cite{Paulus2000,NanoOpticsBook}
\begin{align}
\mathbf{G}(\mathbf{r},\mathbf{r},\omega ) = \frac{i}{8\pi
k^2} \int_0^\infty d k_\rho \frac{k_\rho}{k_{z}} (
\mathbf{M}_0 + \mathbf{M}_r ),
\end{align}
where
\begin{align}
\mathbf{M}_0 =& \left[
\begin{array}{ccc}
2k^2-k_\rho^2 & 0 & 0 \\
0 & 2k^2 -k_\rho^2 & 0 \\
0 & 0 & 2(k^2-k_{z}^2)
\end{array} \right]
\end{align}
and
\begin{align}
\mathbf{M}_r =& \left[
\begin{array}{ccc}
(k^2 r^s - k_{z} r^p) & 0 & 0 \\
0 & (k^2 r^s - k_{z}^2 r^p) & 0 \\
0 & 0 & (2k_\rho^2 r^p)
\end{array} \right] e^{2 i k_{z} z}.
\end{align}
Here $k = \left| {\bf k} \right|$, where $\mathbf{k} = (k_\rho , k_z , k_\phi)$ is the
$\mathbf{k}$-vector in cylindrical coordinates, $z$ is the
distance from the QD to the interface, and $r^s$ ($r^p$) is the
Fresnel reflection coefficient for s-polarized (p-polarized)
light \cite{NanoOpticsBook}. The result of the calculation
for a $\mathrm{GaAs}$-air interface is shown in Fig.~\ref{Figure1_ver17}(c)

We now consider the transition matrix element. Using the fact that
the momentum operator is a differential operator
$(\mathbf{p}=-i\hbar \mathbf{\nabla})$, that the Bloch functions $
u_{c/v,0}(\mathbf{r})$ are orthogonal, which we will describe
below, and that the envelope functions are slowly varying on the
scale of a lattice parameter \cite{Coldren_and_Corzine}, we obtain
\begin{align}
\begin{split}
\left| \langle v | \mathbf{p} | c \rangle \right|^2 \approx&
\left|\langle F_h | F_e \rangle \right|^2 \left| \langle u_{v,0} |
\mathbf{p} | u_{c,0} \rangle \right|^2\\
\equiv& \left|\langle F_h | F_e \rangle \right|^2 \times \frac{m_0 E_p(c)}{2}
.\label{eq:Matrix_Element_Product_Form}
\end{split}
\end{align}
This important result states that the transition matrix element is
given by the product of the electron and hole wavefunction overlap
and the squared Bloch matrix element $\left| \langle u_{v,0} |
\mathbf{p} | u_{c,0} \rangle \right|^2$. The magnitude of the Bloch
matrix element is a material parameter that is expressed in terms
of the Kane energy $E_p(c)$ \cite{Coldren_and_Corzine}. $c$ is the
indium mole fraction in the $\mathrm{In_cGa_{1-c}As}$ alloy. The

In the Kane model \cite{Chuang,Davies} the valence band Bloch
functions are written as linear combinations of the basis
functions $|u_{x}\rangle$, $|u_{y}\rangle$, and $|u_{z}\rangle$
that carry the symmetry properties of p-orbitals. The specific
linear combination depends on the $\mathbf{q}$-vector of the
envelope function. QDs grown by the Stranski-Krastanov technique
are typically flat structures placed on top of a wetting layer,
and quantization along the growth direction $(z)$ is therefore
dominating \cite{Andreev2005}. As a consequence we can set $q_x =
q_y = 0$, which is exact in the limit of a quantum well, leading
to $\mathbf{q}=|\mathbf{q}|\hat{\mathbf{e}}_z$. In this case the
heavy hole Bloch function can be written as $| u_{v,0} \rangle =
|u_{hh}\rangle = \frac{1}{\sqrt{2}}\left(|u_{x}\rangle \pm
i|u_{y}\rangle\right)$. The conduction band Bloch function has
s-symmetry, so the Bloch functions for the valence and conduction
bands are orthogonal as was used above. Furthermore, this means
that the matrix element $\left| \langle u_{v,0} | \mathbf{p} |
u_{c,0} \rangle \right|^2$ is non-zero only for $p_x$ and $p_y$
from which we conclude that the dipole axis of the QD is
perpendicular to the growth direction in agreement with our
experiment.

\subsection{Frequency dependence of the radiative decay rate of quantum dots}
We are now in a position to put together the results of the
previous sections and calculate the frequency dependence of the
spontaneous emission decay rates in a homogenous medium. By insertion of Eq.~(\ref{eq:LDOS_def}) in
Eq.~(\ref{eq:Gamma_rad}) and using the plane wave expression for the field distribution functions we readily obtain
\begin{align}
\begin{split}
\Gamma&_\mathrm{rad}^{hom}(\omega) = \\ & \frac{\pi q^2}{\hbar m_0^2
\epsilon_0} \frac{1}{\epsilon_r V} \left| \langle v | \mathbf{p} | c \rangle \right|^2 \frac{1}{\omega_0} \sum_{\mathbf{\mu}}
|\mathbf{\hat{e}}_\mathbf{p} \cdot
\mathbf{\hat{e}}_\mathbf{\mu}|^2
\delta(\omega-\omega_\mathbf{\mu}).
\end{split}
\end{align}
The sum over all optical modes $\mathbf{\mu}$ can be converted to
an integration over all k-vectors where the dispersion relation
for a homogeneous medium $\omega_{\mu} = k_{\mu} c_0/n$ is used. The sum over polarizations yields a factor of $2$. Using also
Eq.~(\ref{eq:Matrix_Element_Product_Form}) we obtain the important
relation
\begin{align}
\Gamma_\mathrm{rad}^{hom}(\omega) = \frac{n q^2}{6 \pi \hbar m_0
c_0^3 \epsilon_0} E_p(c) \omega \left|\langle F_h (\omega) | F_e
(\omega) \rangle \right|^2\label{eq:Gamma_rad_hom2}.
\end{align}
Here we have indicated explicitly that the envelope wavefunctions
depend on the emission energy since varying the QD size, and
thereby the emission energy, leads to modifications of the
wavefunctions. This effect will be discussed in detail in section
\ref{sec:NumericalResults}. Eq. (\ref{eq:Gamma_rad_hom2}) is the
key result used to interpret the experimental measurements of the
radiative decay rate presented in this paper. It furthermore
allows extracting an experimental value for the overlap between
the electron and hole wavefunctions.

\section{Comparison between experiment and theory for the electron-hole wavefunction overlap\label{sec:NumericalResults}}

\begin{figure}
\includegraphics[width=\columnwidth]{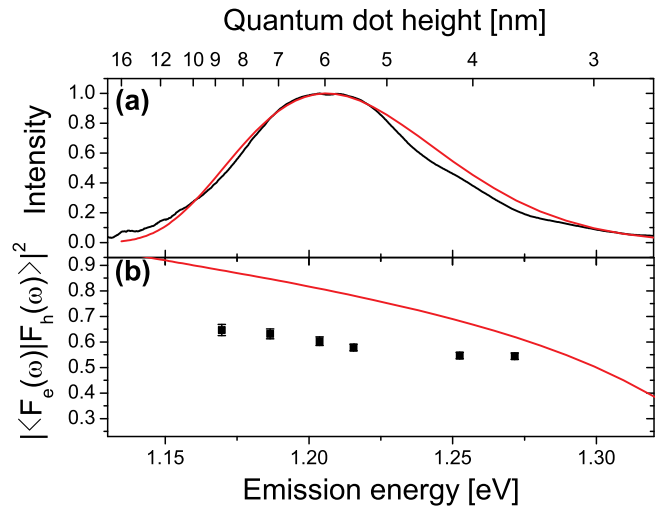}
\caption{(Color online) (a) Measured
spontaneous emission spectrum (black curve, normalized) and
calculated spectrum (red curve) using the height distribution of
the QDs from Fig.~\ref{Figure2_ver17}(b) and implementing strain
model 1. The resulting parameters have been
optimized to fit the spectrum, and we find an aspect ratio of
$1/6$ and an indium mole fraction in the QD of $39\%$. (b)
Measured electron-hole wavefunction overlap (black squares) and
the theoretical calculation (solid red line) using the same
parameters as in (a). \label{Figure11_ver17}}
\end{figure}

\begin{figure}
\includegraphics[width=\columnwidth]{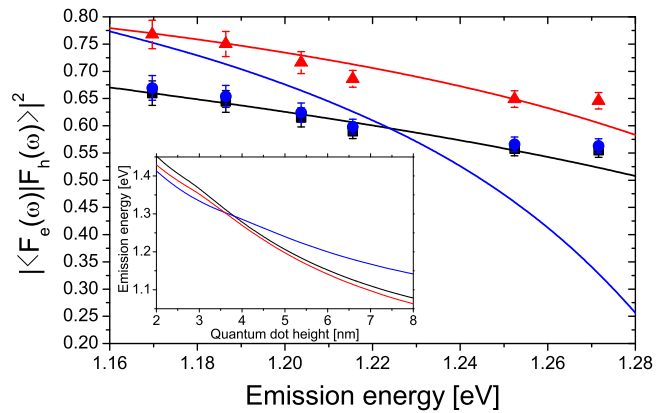}
\caption{(Color online) Calculated energy dependence of the squared electron and
hole wavefunction overlap using the three different strain models
discussed in the text. The QD aspect ratio is $1/2$ and no wetting
layer was included. The curves are calculated for strain model 1
with $c=0.51$ (blue line), strain model 2 with $c=0.95$ (red line)
and strain model 3 with $c=0.46$ (black line). The data points
show the experimentally determined wavefunction overlap for these
three indium mole fractions using the same color coding. Clearly,
strain model 1 does not describe the experimental data well, while
both models 2 and 3 lead to very good agreement. The inset shows
the dependence of the QD emission energy on height for the three
different models using the same color coding as in the main
figure.
\label{Figure12_ver17}}
\end{figure}

\begin{figure}
\includegraphics[width=\columnwidth]{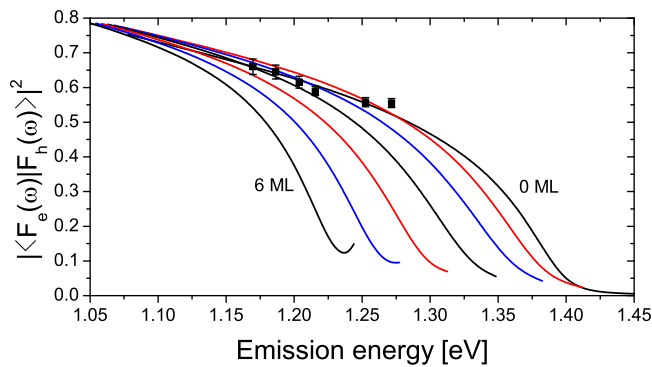}
\caption{(Color online) The effect of the thickness of the wetting layer on the
calculated electron-hole wavefunction overlap when varying between
0 ML and 6 ML in steps of 1 ML. Here we have omitted strain
 and otherwise used the same parameters as in
Fig.~\ref{Figure12_ver17}. The experimental data for $c=0.46$ are
shown as black squares. \label{Figure13_ver17}}
\end{figure}

\begin{figure}
\begin{center}
\includegraphics[width=\columnwidth]{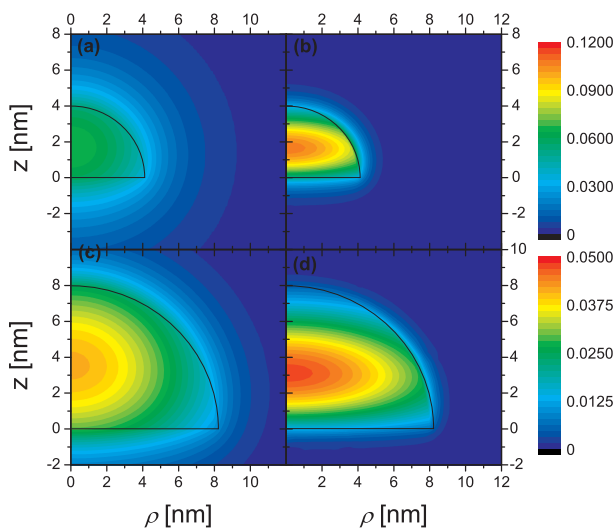}
\caption{(Color online) Color plots in the radial plane of the amplitude of
the wavefunction for electrons and holes. The parameters corresponding to the calculation for
strain model 2 in Fig.~\ref{Figure12_ver17} were used.  (a)
Electron wavefunction for a QD height of $4$~$\mathrm{nm}$. (b)
Hole wavefunction for a QD height of $4$~$\mathrm{nm}$. (c)
Electron wavefunction for a QD height of $8$~$\mathrm{nm}$. (d)
Hole wavefunction for a QD height of $8$~$\mathrm{nm}$.
\label{Figure14_ver17}}
\end{center}
\end{figure}

In this section we calculate the electron-hole wavefunction
overlap and the QD emission spectrum and compare to our
experimental results. We investigate to what extent the QD heights
obtained from AFM measurements of uncapped QDs, cf. Fig.~\ref{Figure2_ver17}, can be used as input to the models.
Furthermore, we systematically test the model with experiment by
varying parameters such as the indium mole fraction, the QD aspect
ratio, the wetting layer thickness, and the applied strain model
within physically realistic boundaries. The model nicely
reproduces the decrease of the electron-hole wavefunction overlap
with energy that we observed experimentally. However, our
investigations lead to the conclusion that further knowledge of QD
composition and more involved QD models would be required in order
to reach full quantitative agreement between experiment and
theory.

We describe the QD in the $(\rho , z)$-plane as one quarter of an
ellipse with a fixed aspect ratio. Here $\rho$ denotes the radial
direction in cylindrical coordinates. The QD geometry is solved
numerically in a large simulation area which spans
$160$~$\mathrm{nm}$ in the $\rho$-direction and $80$~$\mathrm{nm}$
in the $z$-direction ensuring that the proximity of the boundaries
have no effect on the results. Further details on the numerical
procedure is provided in Appendix
\ref{sec:NumericalModelingAppendix}. We use the commercial finite
element software package \emph{COMSOL} with an adaptive mesh to
solve the effective mass equation (Eq.~(\ref{eq:Hamiltonian7})).
For each QD height we obtain the transition energy and given the
height distribution function measured by AFM, c.f.
Fig.~\ref{Figure2_ver17}, we calculate the emission spectrum.

The fact that we observe transitions involving heavy holes
motivates a further investigation of the analogy between QDs and
quantum wells. In particular the effect of strain on the
electronic level structure is well-understood for quantum wells
\cite{Chuang}. Strain has a significant effect on the level
structure also for QDs, but is often discussed in qualitative
terms only due to the mathematical complexity and lack of
experimental input on the exact QD geometry and composition. Thus,
we will in the following attempt to model the strain properties of
the QD similar to the case of a quantum well and compare the
results to experiment. In a quantum well of $\mathrm{InGaAs}$ in
$\mathrm{GaAs}$ the compressive strain in the plane of the quantum
well leads to an expansion in the direction perpendicular to the
plane. This biaxial strain can be decomposed into a hydrostatic
and a shear component. However, as opposed to the case of a quantum well a QD cannot expand freely in the growth direction which suggests that the hydrostatic component may dominate. Therefore, we compare
three different strain models including: 1) hydrostatic and shear
strain, 2) only hydrostatic strain, and 3) no strain. Strain
modifies the band offsets and the valence band effective masses,
which is discussed in further detail in Appendix
\ref{sec:strain_appendix}.

Using the experimental data on both the QD height distribution
function, the emission spectrum, and the frequency dependence of
the wavefunction overlap, we explore experimentally realistic
parameters in order to match our experimental data. The first
approach is to use the measured QD height distribution of Fig.~\ref{Figure2_ver17} and include both shear and hydrostatic strain.
Using optimized parameters corresponding to an aspect ratio of
$1/6$ and an indium mole fraction of $39\%$ we find a very good
agreement with the emission spectrum, as shown in
Fig.~\ref{Figure11_ver17}(a). The electron-hole wavefunction
overlap extracted from the radiative decay rate offers an
additional test of this parameter set. In
Fig.~\ref{Figure11_ver17}(b) the experimentally determined
wavefunction overlap is plotted along with the
resulting theory \cite{Johansen2008ErrorComment}. The theory predicts correctly that the
wavefunction overlap decreases (increases) with emission energy
(QD size), and the mechanism behind this effect is discussed in
detail below. However, a clear systematic deviation between theory
and experiment is observed, which turns out to be the case for all three strain models provided that the model parameters are constrained to optimally reproduce the measured spectrum.
We conclude from this result that the measured height distribution
of uncapped QDs is \emph{not} very useful in determining the
actual confinement volume of the overgrown QDs that the
experiments are performed on. This is due to complex redistribution and intermixing processes
of indium and gallium will occur during growth and subsequent
regrowth \cite{Kegel2000,Liu2000}, which are likely to modify the
QD confinement potential and strain significantly.

Thus abandoning at this point the interpretation of the measured
height distribution, which is closely related to the emission
spectrum, we focus on the electron-hole wavefunction overlap.
Keeping the aspect ratio fixed at a realistic value of $1/2$ and
excluding for the moment the wetting layer, we obtain the overlap
shown in Fig.~\ref{Figure12_ver17}. Here the only free parameter
is the mole fraction of indium $c$ in the QD and the different
strain models are compared. Note that the experimentally
determined wavefunction overlap depends on $c$, since it enters
through the Kane energy, cf. Eq. (\ref{eq:Gamma_rad_hom2}). We
find that including both shear and hydrostatic strain leads to a
disagreement between experiment and theory. This demonstrates that
the strain model developed for quantum wells fails in the case of
QDs, which is sometimes assumed in the literature \cite{Li1996}. Good agreement between experiment and theory can be
obtained when including only hydrostatic strain or no strain at
all for $c=0.95$ and $c=0.46$, respectively. Judging from
experiments available in the literature \cite{Liu2000,Kegel2000}
both these values of $c$ are reasonable for overgrown QDs, so
there is no support for favoring one of the two surviving strain
models. We are led to the conclusion that in particular shear
strain is less significant for QDs compared to quantum wells,
although further microscopic details of QD composition and
geometry would be required for a further investigation of these
issues.

Figure~\ref{Figure13_ver17} investigates the effect of the wetting
layer thickness on the wavefunction overlap. In this case strain
is omitted, the aspect ratio is $1/2$, and the indium mole
fraction $0.46$. We find that for wetting layer thicknesses below
4 monolayer (ML) the wavefunction overlap is only slightly
modified in the emission energy range of interest to the
experiment. For a very thick wetting layer (6 ML), the QD
wavefunction is modified such that the monotonic decrease of the
wavefunction overlap with energy observed for all other
thicknesses does not apply. This behavior can be understood as
follows: for very thick wetting layers and small QDs a significant
part of both the electron and hole wavefunctions are expelled from
the QD giving rise to quantum well-like wetting layer states that
can have an increased mutual overlap. For the QD sample
of the experiment the wetting layer was on the order of 2 ML.

The above discussions point to a number of subtleties associated
with a quantitative comparison between experiment and theory.
These are mainly related to lack of knowledge about intrinsic
properties of the QDs. Notably, the monotonic decrease in the
electron-hole wavefunction overlap with emission energy is found
to be a very general and robust result for a large range of
different parameters. The generality of this result can be
understood in a simple physical picture and is related to the
differences in the electron and heavy hole effective masses. Thus,
for any indium mole fraction, we have from
Eqs.~(\ref{eq:HeavyHoleMassBulk}), (\ref{eq:HeavyHoleMassQDx}),
and (\ref{eq:HeavyHoleMassQDz}) that $m_{hh,b} > m_{e,b}$ and
$m_{hh,x},m_{hh,z} > m_{e}.$ This in turn leads to a smaller Bohr $a_{B_n} = \frac{4\pi \epsilon_0 \epsilon_r \hbar^2}{q^2 m_0 m_n}$
radius for holes than for electrons, i.e. the hole wavefunctions
are more compressible than the electron wavefunctions. Increasing the emission energy corresponds to decreasing
the QD size. The large QDs emitting at small energies have a
relatively large electron-hole wavefunction overlap. Decreasing QD
size (i.e. increasing emission energy) compresses the electron and
hole wavefunctions, and since the quantum confinement effect
influences the electron wavefunction more than the hole wavefunction,
the wavefunction overlap is decreased. This effect is illustrated
in Fig.~\ref{Figure14_ver17} where the calculated electron and
hole wavefunctions for two different QD sizes are plotted. It is
clearly observed that a reduction in the QD height leads to a
compression of the hole envelope wavefunction while the electron
wavefunction extends further into the surrounding $\mathrm{GaAs}$
barrier.

\section{Conclusion\label{sec:conclusion}}

We have presented time-resolved measurements of spontaneous
emission from self-assembled QDs near a semiconductor-air
interface. The interface leads to a modification of the LDOS,
which can be calculated in an exact model without any adjustable
parameters. The excellent agreement between theory and experiment
enables separating radiative and non-radiative decay rates whereby
they can be determined with unprecedented accuracy. We reviewed
the theory behind the experiment by calculating the spontaneous
emission radiative decay rate in a full quantum model in which
spontaneous emission is described by Wigner-Weisskopf theory. The
radiative decay rate is proportional to the projected LDOS, which
was derived using a Green's function formalism.

From our measurements of the radiative decay rate at different
emission energies we extracted the frequency dependence of the
overlap of the electron and hole envelope wavefunctions. The
experimental data were compared to theory by solving numerically
the Schr\"{o}dinger equation for a QD potential including the
effects of shear and hydrostatic strain. From this model the
spontaneous emission spectrum, which is inhomogeneously broadened
due to the different sizes of QDs making up the ensemble, and the
electron-hole wavefunction overlap were derived. An attempt to
model the emission spectrum using the QD height distribution
obtained by AFM on uncapped QDs was unsuccessful leading to the
conclusion that this height distribution does not properly reflect
the microscopic confinement potential of overgrown QDs. Regarding
the frequency dependence of the  electron-hole wavefunction
overlap, we found good agreement between experiment and theory
with reasonable assumptions about QD size, geometry, strain, and
wetting layer thickness assuming purely hydrostatic strain or no
strain at all. In contrast, systematic deviations were found when
including both shear and hydrostatic strain. Although the
numerical model employed here is too simple to reflect the
microscopic details of the QD geometry such as, e.g.,
indium-gallium intermixing, it reflects this simple physical
picture very well. A more detailed comparison between experiment
and theory is limited by the lack of experimental input on the QD
structure at the atomic scale, which would be required to verify
more sophisticated QD models.

We finally discussed how the striking frequency dependence of the
radiative decay rate, and consequently of the electron-hole
wavefunction overlap, can be understood in terms of a very simple
physical picture: The heavy hole wavefunction is more compressible
than the electron wavefunction due to the larger effective mass. A
reduction in the QD size therefore leads to a further localization
of the heavy hole wavefunction, while the electron wavefunction
delocalizes into the surrounding barrier. This leads to a decrease
in wavefunction overlap for increasing emission energy.

Quantitative comparison of the experimental data to theory was
limited by the lack of detailed experimental input about the
microscopic composition of the QDs. Combining the detailed optical
experiments presented here with techniques to extract local
material properties of QDs, e.g., by high resolution transmission
electron microscopy (TEM), will be a very exciting future research
direction that also will pinpoint the need for more involved
theoretical models of the QDs. We believe that the technique
presented here to directly access the light-matter coupling
strength will have important applications regarding proper
design and characterization of solid state quantum photonic devices.

\begin{acknowledgments}
We wish to thank J.~H{\o}jbjerg, A.~Kreiner-M{\o}ller and
J.~E.~Mortensen for valuable work on the numerical model and
C.B.~S{\o}rensen for growth of the semiconductor material. We
gratefully acknowledge the Danish Research Agency for financial
support (projects FNU 272-05-0083, 272-06-0138 and FTP
274-07-0459).
\end{acknowledgments}

\appendix

\section{Influence of strain on excitons confined in quantum dots \label{sec:strain_appendix}}

\begin{figure}
\includegraphics[width=\columnwidth]{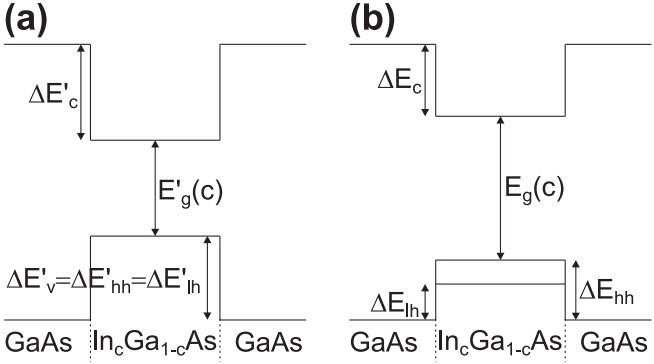}
\caption{Schematic illustration
of the effect of strain on the band energies at
$\mathbf{q}=\mathbf{0}$. (a) No strain and (b) biaxial strain.
Biaxial strain shifts both the conduction band and the heavy hole
valence band closer to the continuum. The net effect is an
increase in the band gap and therefore in the transition energy.
Importantly, the degeneracy of the light and heavy hole states is
lifted and the lowest energy transition involves heavy holes only.
\label{Figure09_ver17}}
\end{figure}

\begin{table*}
\caption{Material parameters for $\mathrm{In_cGa_{1-c}As}$ at
cryogenic temperatures used in this work. CB and VB indicate conduction and valence band parameters respectively.}
\begin{tabular}{lcccc}\hline\hline%{|lc|c|c|c|}\hline\hline
Quantity & & Value for $\mathrm{In_cGa_{1-c}As}$ & Unit & Reference(s)\\
\hline Lattice constant & $a$ & $5.6503+0.4050c$ & \AA\ & \cite{Stier1999}\\
Band gap & $E_g$ & $1.515-1.580c+0.475c^2$ & $\mathrm{eV}$ & \cite{Stier1999}\\
CB effective mass & $m_e$ & $0.0667-0.0419c-0.00254c^2$ & $m_0$ & \cite{Stier1999}\\
Luttinger parameter & $\gamma_1$ & $1/[(1-c)/6.98+c/20.0]$ & & \cite{Stier1999,Vurgaftman2001}\\
Luttinger parameter & $\gamma_2$ & $1/[(1-c)/2.06+c/8.5]$ & & \cite{Stier1999,Vurgaftman2001}\\
Luttinger parameter & $\gamma_3$ & $1/[(1-c)/2.93+c/9.2]$ & & \cite{Stier1999,Vurgaftman2001}\\
CB Hydrostatic def. pot. & $a_c$ & $-8.013 + 2.933c$ & $\mathrm{eV}$ & \cite{Stier1999}\\
VB Hydrostatic def. pot. & $a_v$ & $-1.824+0.024c$ & $\mathrm{eV}$ & \cite{Stier1999}\\
VB Shear def. pot. & $b_v$ & $-2.0+0.2c$ & $\mathrm{eV}$ & \cite{Vurgaftman2001}\\
Elastic stiffness constant & $C_{11}$ & $1221-388.1c$ & $\mathrm{GPa}$ & \cite{Vurgaftman2001}\\
Elastic stiffness constant & $C_{12}$ & $566-113.4c$ &
$\mathrm{GPa}$ & \cite{Vurgaftman2001}\\
Static dielectric constant & $\epsilon_r$ & $13.18+1.42c$ & &
\cite{Stier1999}\\ Kane Energy & $E_p(c)$ & $28.8-7.3c$ &
$\mathrm{eV}$ &
\cite{Coldren_and_Corzine,Vurgaftman2001}\\
 \hline\hline
\end{tabular}
\label{Table1_ver17}
\end{table*}

Strain due to the lattice constant mismatch between
$\mathrm{InAs}$ and $\mathrm{GaAs}$ is responsible for the
formation of QDs during MBE growth in the Stranski-Krastranov
growth mode. This means that the QDs are highly strained and this
has significant impact on the electronic band structure. The
interplay between geometry, chemical composition, and strain is
complicated for QDs. During the growth, diffusion of $\mathrm{In}$
and $\mathrm{Ga}$ takes place, so that the resulting QD will
consist of a significant fraction of $\mathrm{Ga}$ even if it is
grown by pure $\mathrm{InAs}$ \cite{Kegel2000}. Furthermore, it
has been reported that $\mathrm{In}$ is mainly concentrated in an
inverted cone inside the QD giving rise to a complex strain
profile \cite{Liu2000}. Complete knowledge about such complex
details is still lacking, and the purpose here is to introduce
simple strain models and judge their validity by comparing to
experimental data. We consider a lens-shaped QD with a lateral
extension, which is larger than the extension in the growth
direction. For this reason we approximate the strain of a QD by
the model used in the case of a quantum well. This is further
motivated by the fact that the QD is placed on top of a wetting
layer, c.f. Fig.~\ref{Figure10_ver17}. The strain modifies the
band offsets and energy gap of the QD, and the valence band
degeneracy is lifted so that the transition with the lowest energy
involves only heavy holes. This is illustrated in Fig.~\ref{Figure09_ver17}. We will assume that the bulk $\mathrm{GaAs}$
surrounding the QD is unstrained, however only heavy hole bands
are included here as well since the influence of band mixing is
minor in the barrier where the electron and hole wavefunctions are
strongly damped.

One effect of strain is to shift the conduction and valence bands
in energy. The strain describes the compression or expansion of
the crystal lattice, which in general is described by a tensor
$\epsilon_{nm}$. For biaxial strain, which describes the strain at
planar heterojunctions, only the diagonal elements are relevant.
We consider a thin layer of $\mathrm{In_cGa_{1-c}As}$ with lattice
constant $a_{\mathrm{QD}}$ embedded in a $\mathrm{GaAs}$ barrier
with lattice constant $a_{b}$. We assume that all strain is
incorporated in the $\mathrm{In_cGa_{1-c}As}$ layer and since
$a_{\mathrm{QD}}>a_{b}$ the strain will be compressive. We have
\cite{Chuang}
\begin{align}
\epsilon_{xx} =\epsilon_{yy} =
\frac{a_b-a_\mathrm{QD}}{a_\mathrm{QD}} \hspace{10ex}
\epsilon_{zz} = -\frac{2C_{12}}{C_{11}}\epsilon_{xx},
\end{align}
where $C_{11}$ and $C_{12}$ are the elastic stiffness constants
(matrix elements of the stiffness tensor). These strain components
lead to a change in band structure and thus a modification of band
energies and effective masses, as described by the Pikus-Bir
strain model \cite{Chuang}. Biaxial strain can be decomposed into components of hydrostatic and shear strain. The hydrostatic
compressive strain of the QD leads to a decrease of the band
offsets exactly as for any hydrostatic compressive strain
resulting from, e.g., a decrease of the temperature. For a thin
strained epitaxial layer, the energy can be lowered by
compensating the in-plane compressive strain by an expansion in
the $z$-direction (shear strain). We obtain the following heavy hole
valence/conduction band offsets \cite{Chuang}
\begin{align}
\Delta E_{hh} &= \Delta E'_{v} - P_\epsilon - Q_\epsilon, \label{eq_DeltaE_hh}\\
\Delta E_{c} &= \Delta E'_{c} - R_\epsilon,
\end{align}
where
\begin{align}
P_\epsilon &= a_v(\epsilon_{xx}+\epsilon_{yy}+\epsilon_{zz}),\\
Q_\epsilon &= -\frac{b_v}{2}(\epsilon_{xx}+\epsilon_{yy}-2\epsilon_{zz}),\label{eq:Qstrain}\\
R_\epsilon &= a_c(\epsilon_{xx}+\epsilon_{yy}+\epsilon_{zz}),
\end{align}
and we have introduced a number of quantities defined in Table
\ref{Table1_ver17}. $\Delta E'_{c}$ ($\Delta E'_{v}$) is the
unstrained conduction (valence) band offset which constitute
$60\%$ ($40\%$) of the band gap difference between bulk and QD, so
that, e.g., $\Delta E'_{c} = 0.6(E_g'(0)-E_g'(c))$. The band gap
of the strained QD is given by
\begin{align}
E_g(c) = E_g'(c) + P_\epsilon + Q_\epsilon + R_\epsilon,
\end{align}
where $E_g'(c)$ is the unstrained band gap of the material. These
effects on the band structure are illustrated in
Fig.~\ref{Figure09_ver17}.

Another important consequence of strain is that the effective
heavy hole mass becomes highly anisotropic. In contrast the
effective electron mass is not modified considerably since the
conduction band is much more energetically isolated than the
valence bands \cite{Chuang,Vurgaftman2001}. In our experiments the
growth direction $(z)$ is parallel to the $[001]$ crystal axis,
which is perpendicular to the wetting layer plane $(x,y)$ oriented
along $[110]$ and $[1\bar10]$ directions. We will use $m_z =
m_\perp = m_{[001]}$ and $m_x = m_\parallel = m_{[110]}$.
Furthermore, since the crystal structure along the $[110]$ and
$[1\bar{1}0]$ axes are identical apart from a series of rotations,
we have that $m_x = m_y$.

The heavy hole masses for unstrained bulk
$\mathrm{In_cGa_{1-c}As}$ are given by
\begin{align}
m'_{hh,\parallel,b} &= \frac{2}{2\gamma_1-\gamma_2-3\gamma_3},\label{eq_heavy_hole_mass_unstrained1}\\
m'_{hh,\perp,b} &= \frac{1}{\gamma_1 -
2\gamma_2}\label{eq_heavy_hole_mass_unstrained2},
\end{align}
where $\gamma_1$, $\gamma_2$ and $\gamma_3$ are Luttinger
parameters \cite{Vurgaftman2001}. These are listed along with all
other relevant QD material parameters for this work in
Table~\ref{Table1_ver17}. For $\mathrm{In_cGa_{1-c}As}$, we have
$\gamma_2 \approx \gamma_3$ allowing for the axial approximation.
Here $\gamma_2$ and $\gamma_3$ are replaced by their average value
$\bar{\gamma}=(\gamma_2+\gamma_3)/2$, leading to an isotropic
heavy hole mass valid for unstrained bulk semiconductors
\cite{Chow_and_Koch}
\begin{align}
m'_{hh,\parallel,b} = m_{hh,\perp,b} = \frac{1}{\gamma_1 -
2\bar{\gamma}}. \label{eq:HeavyHoleMassBulk}
\end{align}
These bulk effective masses are used to describe the
$\mathrm{GaAs}$ barrier material surrounding the QDs. The strained
heavy hole effective masses in the directions parallel with and
perpendicular to the wetting layer plane are given by
\cite{Chow_and_Koch,Coldren_and_Corzine,Chuang}
\begin{align}
m_{hh,\parallel,\mathrm{QD}} &= \frac{1}{\gamma_1+\bar{\gamma}},\label{eq:HeavyHoleMassQDx}\\
m_{hh,\perp,\mathrm{QD}} &=  \frac{1}{\gamma_1 -
2\bar{\gamma}}\label{eq:HeavyHoleMassQDz},
\end{align}
showing that the parallel component is strongly
modified by strain.

\section{Numerical modeling of envelope wavefunctions\label{sec:NumericalModelingAppendix}}
We solve the effective mass equation for electrons in the
conduction band and holes in the valence band in order to
calculate the overlap of the envelope wavefunctions. The effective
mass equations describing the electron and the hole have the same
form, but the effective hole mass anisotropy must be taken into
account. We therefore consider the anisotropic valence band
problem which has the isotropic conduction band problem as a
special case. For both electrons and holes the effective mass
depends on position due to the different effective masses in the
QD and in the surrounding crystal matrix.

Using cylindrical coordinates $(\rho,\phi,z)$ and the axial approximation, the kinetic term in Eq.~(\ref{eq:TheEffectiveMassHamiltonian}) reads
\begin{align}
\begin{split}
H_\mathrm{kin} =&
-\frac{\hbar^2}{2 m_0} \bigg(
\frac{1}{\rho}\frac{\partial}{\partial \rho} \bigg(
\frac{\rho}{m_{n\parallel}(\rho,z)} \frac{\partial}{\partial \rho}
\bigg)
+\frac{1}{m_{n\parallel}(\rho,z) \rho^2}
\frac{\partial^2}{\partial \phi^2}\\
&+ \frac{\partial}{\partial z}
\bigg( \frac{1}{m_{n\perp}(\rho,z)} \frac{\partial}{\partial z}
\bigg)\bigg).
\end{split}
\end{align}
Using separation of variables
$F(\mathbf{r})=R(\rho,z)\Phi(\phi)$ the effective mass
Schr\"{o}dinger equation  can be written as
\begin{align}
\begin{split}
\frac{-1}{\Phi(\phi)} \frac{\partial^2}{\partial \phi^2} \Phi(\phi)
=& \frac{\rho}{R(\rho,z)} \frac{\partial}{\partial \rho} R(\rho,z)\\
&+ \frac{m_{n\parallel}(\rho,z) \rho^2}{R(\rho,z)} \frac{\partial}{\partial \rho} \left( \frac{1}{m_{n\parallel}(\rho,z)} \frac{\partial}{\partial \rho} R(\rho,z) \right)\\
&+ \frac{m_{n\parallel}(\rho,z) \rho^2}{R(\rho,z)} \frac{\partial}{\partial z} \left( \frac{1}{m_{n\perp}(\rho,z)} \frac{\partial}{\partial z} R(\rho,z) \right)\\
&+ \frac{2m_0 m_{n\parallel}(\rho,z) \rho^2}{\hbar^2} (V(\rho,z)-E).
\label{eq_Schrodinger_cylindrical1}
\end{split}
\end{align}
The left and right hand sides of this equation are independent and
they must therefore equal a constant, i.e. $\frac{1}{\Phi(\phi)}
\frac{\partial^2}{\partial \phi^2} \Phi(\phi) = -l^2$. The
solution of this equation is $\Phi(\phi) = c_1 cos(l\phi) + c_2
sin (l\phi)$ which by the boundary condition $\Phi(0)=\Phi(2\pi)$
implies that $l$ must be an integer. We are considering the ground
state transition and therefore take $l=0$. This leaves an equation
describing the electronic motion in the $(\rho,z)$-plane.

Eq.~(\ref{eq_Schrodinger_cylindrical1}) is solved numerically
after being reduced to a dimensionless form in order to avoid
numerical issues related to the very small factors
($\propto\hbar^2$) appearing in this equation. We define the new
dimensionless quantities through $\rho = k_\rho \tilde\rho$, $z =
k_z \tilde{z}$, $R(\rho,z) = k_R \tilde{R}(\tilde\rho,\tilde{z})$,
$V(\rho,z) = V_0 \tilde{V}(\tilde\rho,\tilde{z})$, $E = V_0
\tilde{E}$, and $V_0 = \hbar^2/(2m_0 k_\rho^2),$ and furthermore
take $k_\rho = k_z = 1$~$\mathrm{nm}$ so that all spatial
dimensions are measured in units of nanometer. By this
transformation we obtain
\begin{align}
\begin{split}
\frac{1}{m_\parallel(\tilde\rho,\tilde{z}) \tilde\rho} \frac{\partial}{\partial \tilde{\rho}} \tilde{R}(\tilde\rho,\tilde{z})
+\frac{\partial}{\partial \tilde{\rho}} \left( \frac{1}{m_\parallel(\tilde\rho,\tilde{z})} \frac{\partial}{\partial \tilde{\rho}} \tilde{R}(\tilde\rho,\tilde{z}) \right)\\
+\frac{\partial}{\partial \tilde{z}} \left( \frac{1}{m_\perp(\tilde\rho,\tilde{z})} \frac{\partial}{\partial \tilde{z}} \tilde{R}(\tilde\rho,\tilde{z}) \right)\\
+\left(\tilde{V}(\tilde\rho,\tilde{z}) -
\frac{l^2}{m_\parallel(\tilde\rho,\tilde{z})
\tilde{\rho}^2}\right)\tilde{R}(\tilde\rho,\tilde{z}) =\tilde{E}
\tilde{R}(\tilde\rho,\tilde{z}),\label{eq:Hamiltonian7}
\end{split}
\end{align}
which is solved numerically using a finite element method.

\begin{thebibliography}{47}
\bibitem[{Joh()}]{Johansen2008ErrorComment}
\bibinfo{howpublished}{Note that in Fig.3.B of Ref. [14] the theoretical curve
  plotted is mistakenly $\left|\langle F_h (\omega) | F_e (\omega) \rangle
  \right|$ rather than $\left|\langle F_h (\omega) | F_e (\omega) \rangle
  \right|^2$.}
\end{thebibliography}
\end{document}